\documentclass{article}
\usepackage[utf8]{inputenc}
\usepackage{hyperref}       
\hypersetup{
  colorlinks   = true, 
  urlcolor     = green, 
  linkcolor    = red, 
  citecolor   = green 
}
\usepackage{url}            
\usepackage{booktabs}       
\usepackage{amsfonts}       
\usepackage{nicefrac}       
\usepackage{microtype}      
\usepackage{subfigure} 

\usepackage{framed,multirow}

\usepackage{amsmath}
\usepackage{algorithm}
\usepackage{algpseudocode}
\usepackage{graphicx}

\usepackage{amssymb}
\usepackage{latexsym}
\usepackage{bm}
\usepackage{amssymb}
\usepackage{array}
\usepackage{multirow}
\usepackage[english]{babel}
\usepackage[numbers]{natbib}
\usepackage{footnote}
\usepackage{tabularx}
\makesavenoteenv{tabular}
\makesavenoteenv{table}
\usepackage{cleveref}
\crefname{equation}{Eq.}{Eqs.}
\crefname{table}{Table}{Tables}
\crefname{figure}{Fig.}{Figs.}
\crefname{section}{Section}{Secs.}
\crefname{subsection}{Section}{Secs.}
\Crefname{figure}{Fig.}{Figs.}
\Crefname{Algorithm}{Algorihtm}{Algorihtm}
\usepackage{xcolor}
\definecolor{newcolor}{rgb}{.8,.349,.1}

\title{Merging Virtual and Real Environments for Visualizing Seismic Hazards and Risk}
\author{Hamed Nikbakht}
\date{October 2016}

\begin{document}

\maketitle
\begin{abstract}
Earthquake research in the last few decades has led to considerable advances in seismic hazard and risk modeling across academia, industry, and government \citep{Nikbakht2011loading,Nikbakht2013comparison, nikbakht2019HMCMC}. Technological advances such as high performance computing and visualization can further facilitate earthquake hazard and risk research. This work utilizes the CAVE of the Marquette Visualization Laboratory to visualize seismic hazards and risk by integrating hazard characterization, structural modeling, and emergency response \citep{nikbakht2015SSA}. Building upon the framework of performance-based earthquake engineering, site-specific ground motions, which link seismic hazards to structural responses, serve as loading inputs to structural models. The resulting structural responses can then be translated into damage states of building elements in the immediate room environment based on fragility functions. To illustrate, we display a map of the Los Angeles region with ground motions for the Mw7.8 ShakeOut scenario \cite{graves2010broadband,graves2011shakeout}, create a virtual room in a residential building subjected to such earthquake shaking, and simulate emergency response in this immersive environment. The illustrative visualization can be extended to various scenarios and help communicate site- and structure-specific hazards and risk to the general public.
\end{abstract}
\section{Introduction}
\begin{figure}[t!]
 \centering
 \includegraphics[width=0.8\linewidth]{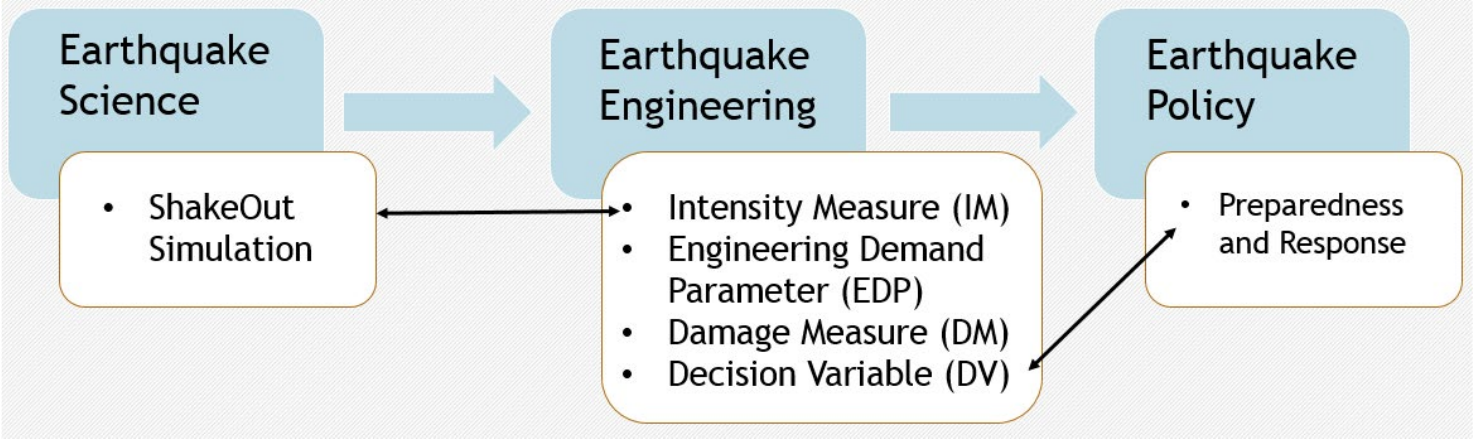}
 \caption{Unifying the ShakeOut scenario with the PBEE framework}
 \label{fig:1}
 \end{figure}

This study aims to visualize emergency response under extreme motions. The “Visualized ShakeOut” completes the cycle of “Rupture to Rafters to Response”. The Shakeout simulation is the hypothetical Mw 7.8 earthquake on the southern San Andreas Fault. The Great ShakeOut Earthquake Drills attracted over 26.5 million participants in 2014 worldwide. To this end, we exploit the 3D, virtual reality and simulation projection technologies and utilizing the CAVE system of the MARquette Visualization Lab (MARVL). 
Furthermore, we implement the Performance Based Earthquake Engineering (PBEE) framework to link the earthquake science, engineering and policy and to visualize emergency preparedness \citep{deierlein2004overview}. The key points in this approach are virtual shaking and merging the virtual and real environments which facilitate earthquake preparedness and response in the CAVE and beyond (\Cref{fig:1}).
\section{Study Framework and Virtual Reality Setup}
\begin{figure}[h]
 \centering
 \includegraphics[width=0.7\linewidth]{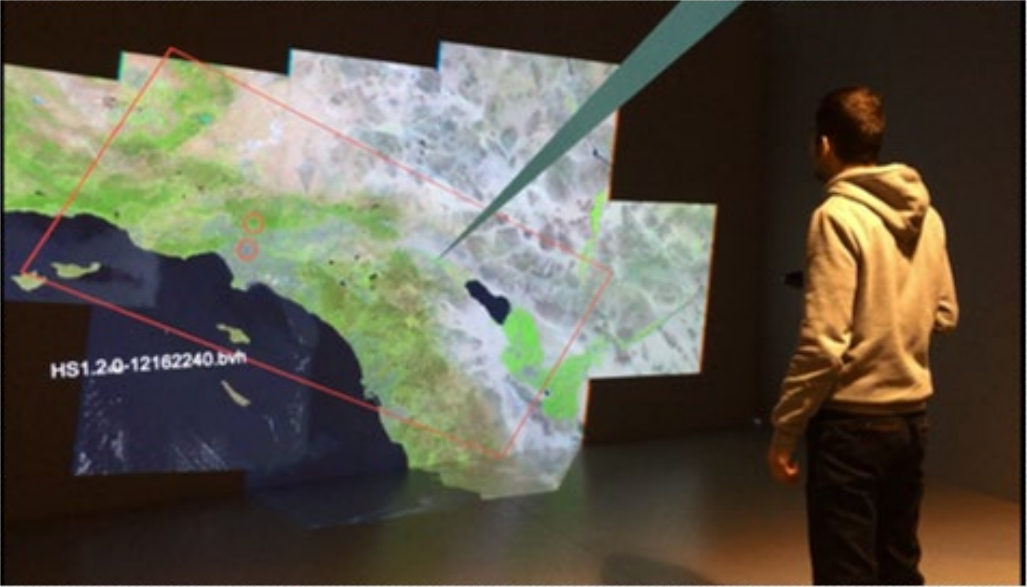}
 \caption{Site map}
 \label{fig:2}
 \end{figure}

In this work, a location marker is placed on a regional map of Southern California. By changing our location on the map, site-specific shaking effects can be simulated on the virtual room (\Cref{fig:2}).\par
\begin{figure}[h]
 \centering
 \includegraphics[width=0.99\linewidth]{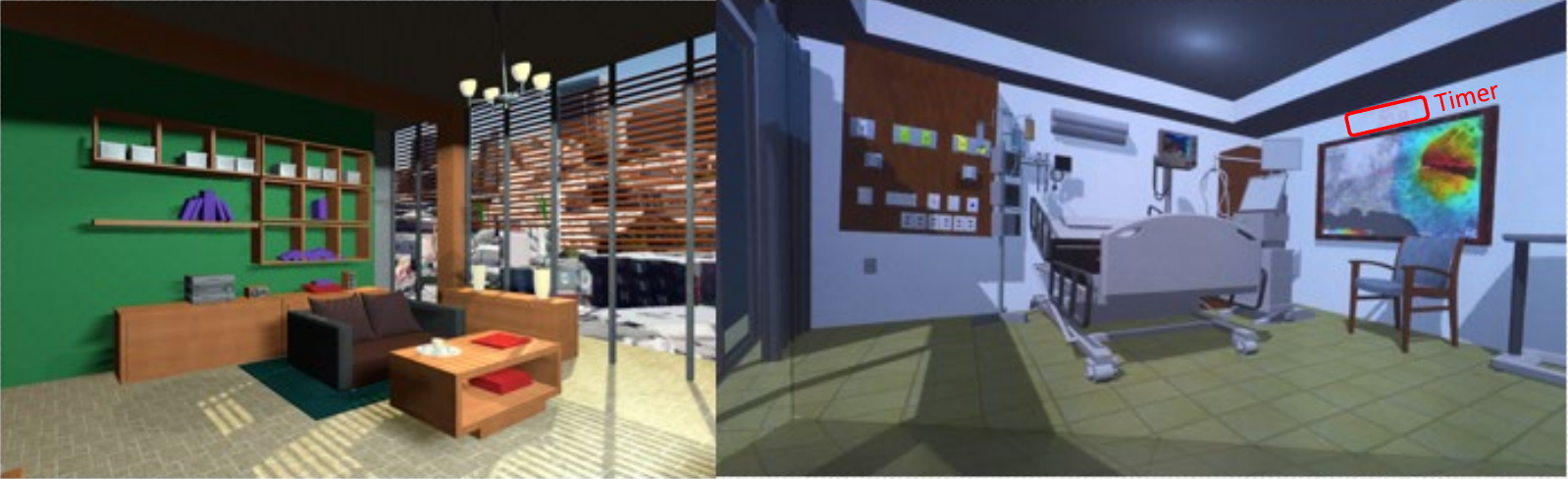}
 \caption{Virtual residential room and hospital emergency room}
 \label{fig:3}
 \end{figure}

For the purposes of demonstration, two virtual rooms - a residence and a hospital room - in downtown Los Angeles are presented here. Additionally, a count-down timer has been assigned to show how much time one has for preparedness before the earthquake strikes the simulated location (\Cref{fig:3}). We thank Dr. Rob Graves for sharing the ShakeOut simulated ground motions for the southern San Andrea’s fault \cite{graves2010broadband,graves2011shakeout}. The ShakeOut map was provided by the U.S. Geological Survey and the Southern California Earthquake Center.\par
 
\begin{figure}[h]
 \centering
 \includegraphics[width=0.7\linewidth]{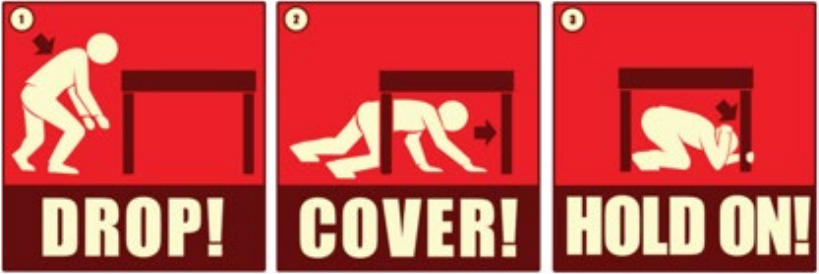}
 \caption{Courtesy of \url{https://www.shakeout.org/}}
 \label{fig:4}
 \end{figure}

To demonstrate the earthquake preparedness actions, we present “Drop, Cover, and Hold On", which are recommended by the Great ShakeOut Earthquake Drills (\url{https://www.shakeout.org/}) as appropriate actions to reduce injury and death during earthquakes (\Cref{fig:4}). These simulations represent an excellent teaching aid regarding scenario-specific emergency response and the potential value of earthquake early warning systems.\par
\begin{figure}[h]
 \centering
 \includegraphics[width=0.8\linewidth]{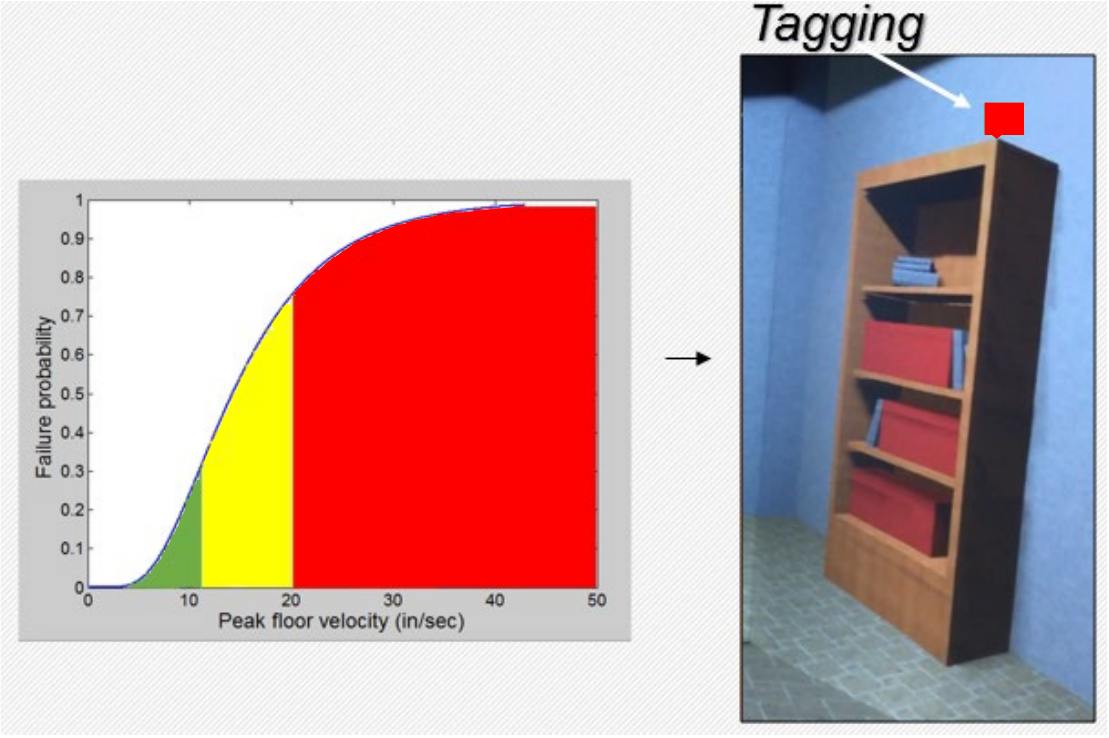}
 \caption{Left figure: Based on empirical data (fragility function of bookcase, 4 shelves, unanchored laterally), Right figure: Based on physical properties of Unity game engine}
 \label{fig:6}
 \end{figure}

In most strong earthquakes, non-structural damage in buildings has been identified as a significant source of casualties and losses. To estimate the physical damage in each visualized model, the Performance Assessment Calculation Tool (PACT), provided by the Applied Technology Council (ATC) \citep{applied2009quantification,applied2012guidelines}, is used to perform the probabilistic computations of losses in non-structural components through the use of fragility functions (\Cref{fig:6}).\par
\begin{figure}[h]
 \centering
 \includegraphics[width=0.6\linewidth]{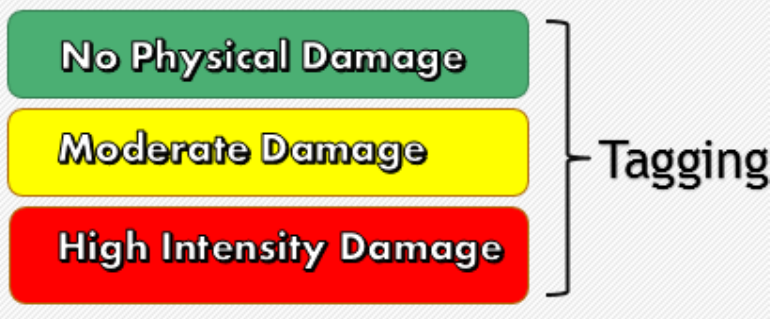}
 \caption{Introducing the Tagging idea}
 \label{fig:5}
 \end{figure}

To demonstrate the simulation of damage measures (the third phase in the PBEE framework), Palm Springs, an additional near-fault site located 2.80 km from the San Andrea’s fault is selected. The damage measure associated with the bookcase is indicated through the use of tagging.  The tags are color-coded to portray the damage level. Green implies little or no physical damage while yellow indicates moderate damage and red indicates high intensity damage on a component (\Cref{fig:5}). This tagging technique could also be applied to structural components. In this case, the final damage level of the bookcase is represented by a red tag (\Cref{fig:6}).\par
 
Furthermore, tagging can also serve as a decision variable, the last phase of performance-based earthquake engineering, because it indicates the expected damage states of each non-structural and structural component in a simulation to aid decision and response.\par
Finally, we present the merged ShakeOut out and visualization demonstrating how the ShakeOut audio changes with shaking intensity (Applying ShakeOut audio as the baseline sound and modulating volume according to shaking intensity).\par
\section{Conclusions}
This work utilizes the CAVE of the Marquette Visualization Laboratory to visualize seismic hazards and risk by integrating hazard characterization, structural modeling, and emergency response. We have shown that the technological advances such as high performance computing and visualization can further facilitate earthquake hazard and risk research. The illustrative visualization can be extended to various scenarios and help communicate site- and structure-specific hazards and risk to the general public.
\section{Acknowledgments}
At the end, I would like to thanks USGS and Dr. Rob Graves for providing simulated ground motion of ShakeOut earthquake scenario, the Southern California Earthquake Center (SCEC) for providing Shakout map and the Applied Technology Council (ATC) for the PACT fragility data. I would also like to show my gratitude to Dr. Ting Lin, my former research advisor, Mr. Christopher Larkee, visualization technology specialist and Dr. John LaDisa, director of the Marquette visualization lab for their contribution and support regarding to modeling the virtual rooms and visualization during this research.
\section*{REFERENCES}
\renewcommand\refname{}
\vspace*{-0.5cm}
\bibliographystyle{unsrt}
\bibliography{Ref}  

\end{document}